\documentclass[aps,pra,floatfix,noshowkeys,epsfig,graphics,natbib]{revtex4}

\usepackage{graphicx}
\usepackage{amsmath}
\usepackage{amsfonts}
\usepackage{amssymb}
\usepackage{comment}
\usepackage{xcolor}
\usepackage{kotex}

\usepackage{amssymb}
\usepackage{amsmath}
\usepackage{epsfig}
\usepackage{graphicx}
\usepackage{slashed}
\usepackage{xcolor,color}
\usepackage{ulem}
\usepackage{subfigure}
\usepackage[toc,page]{appendix}
\usepackage[makeroom]{cancel}
\usepackage{tikz}
\usetikzlibrary{shapes.geometric,positioning,decorations.pathreplacing,decorations.pathmorphing,patterns,decorations.markings}
\usetikzlibrary{scopes,intersections}

\makeatletter

\@addtoreset{equation}{section}
\makeatother
\def\be{\begin{equation}}
\def\ee{\end{equation}}
\def\bea{\begin{eqnarray}}
\def\eea{\end{eqnarray}}
\def\bgn{\begin{align}}
\def\egn{\end{align}}

\def\({\left(}
\def\){\right)}
\def\<{\left<}
\def\>{\right>}

\def\({\left(}
\def\){\right)}
\def\<{\left<}
\def\>{\right>}
\def\!{\right|}
\def\|{\left|}

\def\[{\left[}
\def\]{\right]}

\def\+{\bar}

\usepackage{hyperref}
\hypersetup{colorlinks,%
  citecolor=blue,%
  linkcolor=blue,%
  pdftex}

\newcommand{\newc}{\newcommand}
\newc{\beq}    {\begin{equation}}
\newc{\eeq}    {\end{equation}}

\newc{\beqa}    {\begin{eqnarray}}
\newc{\eeqa}    {\end{eqnarray}}
\newc{\bs}    {\section}
\newc{\no}    {\\ \nonumber}

\def\mnras{{ Mon. Not. Roy. Astron. Soc.  }}

\newc{\st}    {\stackrel}

\begin{document}

\title{
Little Red Dots and Supermassive Black Hole Seed Formation in Ultralight Dark Matter Halos}

\author{Dongsu Bak}
\email{dsbak@uos.ac.kr}
\affiliation{  Physics Department,
University of Seoul, Seoul, 02504, Korea}
\author{Jae-Weon Lee}
\email{scikid@jwu.ac.kr}
\affiliation{Department of Electrical and Electronic Engineering, Jungwon University, 85 Munmu-ro, Goesan-eup, Goesan-gun, Chungcheongbuk-do, 28024, Korea.}

\begin{abstract}
We explore a possible mechanism for the formation of supermassive black hole (SMBH) seeds at the centers of ultralight dark matter (ULDM) halos in the early Universe.
We investigate the conditions under which high-redshift baryonic gas, strongly confined by central solitonic cores of the halos, undergoes direct and monolithic collapse. 
The solitonic core sets characteristic mass and length scales for the confined baryons. Once the confined gas becomes self-gravitating, rapid inflow and shock heating may drive it into a high-temperature and high-density regime favorable for suppressing molecular cooling, without requiring a strong external UV background.
We present semi-analytic scaling relations for the halo mass, soliton mass, baryonic core radius, and characteristic thermodynamic state of the gas, parametrizing the possible effects of baryonic contraction.
These relations provide order-of-magnitude estimates of the characteristic range of SMBH seed masses as a function of redshift.
In this framework, pristine gas clouds satisfying the adopted thermal criterion may avoid efficient fragmentation and undergo rapid central collapse, potentially forming massive black hole seeds with characteristic masses of order $ 10^5 M_\odot$, while systems below the threshold may form compact star clusters instead.
The ULDM particle mass required to reproduce the inferred seed mass scale, $m \simeq O(10^{-22}){\rm eV}$, lies in a range
 favored by galactic-scale observations, suggesting a possible connection between the characteristic scales of galactic cores and early SMBH seeds. 
 Our estimates indicate that favorable conditions for SMBH seed formation may arise at redshifts $z \gtrsim 10$.
 Such conditions may be relevant to the young SMBHs inferred in some little red dots, which appear to be embedded in compact, dense, ionized gas.
\end{abstract}

\maketitle

\section{Introduction}

The detection of bright quasars less than a billion years after the Big Bang indicates that supermassive black holes (SMBHs) with masses up to $\sim 10^9 M_{\odot}$ already existed at redshifts $z \gtrsim 6$, posing a serious challenge to models of black hole seeding and early growth~\cite{Volonteri2010,Inayoshi2020}. In  stellar-remnant seed scenarios, maintaining nearly continuous and efficient accretion is difficult in  shallow potential wells
at high redshift, which motivates massive-seed scenarios where black holes are born with initial masses of $\sim 10^{4}$--$10^{6}M_{\odot}$
~\cite{Volonteri2010,LatifFerrara2016,Inayoshi2020}. 

 The direct-collapse black hole (DCBH) scenario assumes that a primordial, metal-free gas cloud undergoes an almost monolithic collapse to form a supermassive star or a quasi-star that subsequently  gives rise to a massive black hole seed~\cite{BrommLoeb2003,BegelmanVolonteriRees2006}. 
The key requirement for this scenario is the suppression of fragmentation, which could be achieved by maintaining the gas temperature  at $\sim 10^{4} \, {\rm K}$ and preventing efficient molecular hydrogen cooling, often via a strong Lyman-Werner UV background or dynamical heating~\cite{ReganHaehnelt2009,ShangBryanHaiman2010,VisbalHaimanBryan2014,Inayoshi2020}.
Recent observations of little red dots (LRDs)~\cite{Rusakov2026LittleRedDots} suggest the presence of relatively low-mass, rapidly growing SMBHs that may represent an early stage of massive black-hole growth in compact, hot, ionized gas clouds in the early Universe.
It is noteworthy that the lower end of the inferred black-hole mass range in LRDs is within roughly an order of magnitude of the characteristic mass scale associated with some galactic dark-matter cores
$\sim 10^6 M_\odot$~\cite{Strigari:2008ib}.
It is therefore tempting to speculate that these two mass scales are physically connected.

One possible explanation for the characteristic mass scale of galactic cores~\cite{Lee:2015cos}
is ultralight dark matter (ULDM)~\cite{1983PhLB..122..221B,Sin:1992bg,Lee:1995af,Matos:1998vk,Hu:2000ke,B_hmer_2007,Chavanis_2012,Hui:2016ltb,Matos_2024,Ferreira2021,2016PhR...643....1M}, which has been extensively studied as an alternative to cold dark matter (CDM) and may alleviate small-scale issues of CDM such as the cusp problem.
Several studies have investigated the possible connection and dynamical interplay between SMBHs and ULDM solitonic cores at galactic centers~\cite{Lee:2015yws,koo2024final,Brax:2019npi}.
In this model, dark matter consists of extremely light scalar particles 
with mass
$m\simeq 10^{-22}eV$ whose macroscopic de Broglie wavelength leads to quantum pressure supported cores (solitons) embedded in larger halos \cite{SchiveChiuehBroadhurst2014,Hui:2016ltb,Ferreira2021}. These solitonic cores typically
have masses $M_{sol}\gtrsim 10^{6} M_\odot$
and
provide  deep, centrally flat gravitational potential wells that can significantly boost baryonic collapse at high redshift~\cite{PhysRevLett.134.051402},
 while also helping to alleviate the final parsec problem~\cite{koo2024final,PhysRevD.110.023517}.

In this paper, we explore  ULDM-assisted SMBH seed formation in which the solitonic core acts as a deep,
centrally flat
potential well  for direct collapse
of baryonic clouds.
The key idea is that
the soliton defines the characteristic spatial scale of a compact baryonic reservoir. Once this core becomes self-gravitating, baryonic infall and shock heating can drive the gas toward the high-temperature, high-density regime in which molecular cooling becomes subdominant, even without invoking an external UV background.
This study provides an order-of-magnitude theoretical estimate. A complete treatment would require radiation-hydrodynamic simulations and is beyond the scope of the present work.

We develop semi-analytic scalings for (i) the characteristic halo and soliton properties as functions of the ULDM mass $m$ and redshift $z$, and (ii) the thermodynamic state of baryons collapsing inside the soliton potential. 
We investigate whether ULDM solitonic cores can generate the thermodynamic conditions required for DCBH formation, as considered in Ref. \cite{Inayoshi_2012}.
By applying the zone-of-no-return criterion for the suppression of $H_2$ cooling~\cite{Inayoshi_2012}, we identify the parameter regime in which fragmentation is avoided and global collapse can occur~\cite{Omukai2001,BrommLoeb2003,LatifFerrara2016,Inayoshi2020}.
In this way, the ULDM soliton may provide a   gravitational route to sustained high inflow rates and elevated temperatures, enabling the formation of supermassive stars with characteristic masses $\sim 10^{5}M_{\odot}$ that can rapidly evolve into SMBH seeds consistent with high-redshift quasar and
LRD observations \cite{BegelmanVolonteriRees2006,Volonteri2010,Inayoshi2020,bhattacharya2025darkrecipegiantspopulation}.

Section II summarizes the ULDM quantum Jeans scales for halos and the soliton-halo relations used in this work.
Section III derives the baryonic core properties inside ULDM solitons and applies the zone-of-no-return criterion for $H_2$ cooling suppression to determine the conditions for suppressing fragmentation
and the resulting seed mass scale.
Section IV discusses the broader implications of our model for early structure formation, including its physical interpretation of LRDs, as well as the observational consequences and limitations of the scenario.

\section{Jeans length and halo mass function of Ultralight dark matter}

In this section, we present the characteristic scales that determine early halos and the corresponding central solitons. The quantum Jeans length scale is given by~\cite{Hu:2000ke,Hui:2016ltb}
\be
 \lambda_J(z) = \frac{\pi^{3/4}\hbar^{1/2}}{\big( G \rho(z)\big)^{1/4} m^{1/2}}=
9.217\times 10^4\, {\left({1+z}\right)^{-3/4}}{m^{-1/2}_{22}}\left(\frac{0.27}{\Omega_{\rm dm}}\right)^{1/4} \left(\frac{0.7}{h}\right)^{1/2}  {\rm pc},
\ee
where  $m_{22}\equiv m/10^{-22}eV$, $h$ is the reduced Hubble constant, and $\Omega_{\rm dm}$ is the dark matter density parameter~\cite{Adame_2025}.
The dark matter density 
at redshift $z$ is
\be
\rho_{ dm}(z)=\rho_{ dm}(0)\,(1+z)^3.
\ee
The quantum Jeans mass scale is then defined by \cite{Hui:2016ltb,Lee:2015cos,Khlopov:1985fch}
\be
M_{J}(z) =\frac{4\pi}{3} \rho_{dm}(z) \,\lambda_J^3(z)=1.204\times 10^8 \left(\frac{1+z}{m^2_{22}}\right)^{3/4}\left(\frac{\Omega_{\rm dm}}{0.27}\right)^{1/4} \left(\frac{h}{0.7}\right)^{1/2}  M_\odot,
\ee
which suggests that ULDM halos near and above this characteristic scale may provide sufficiently deep potential wells to aid baryonic confinement at high redshift.
In the simplified treatment adopted here, we use $M_J$ as a characteristic lower collapse scale and restrict attention to halos with $M_h \ge M_J$.
We focus on the characteristic mass interval $M_J \lesssim M_h \lesssim M_{\rm upper}$, where the lower scale represents the adopted ULDM collapse threshold and the upper scale
$M_{\rm upper}$ is defined statistically and
reflects the observed number density of galactic halos.

We shall  assume $\Omega_{\rm dm}=0.27$ and $h=0.7$ in this paper. Then,
the dark matter density at redshift $z$ is  given by
\be
\rho_{dm}(z)=
2.485\times 10^{-30}
\,(1+z)^3 {\rm g/cm}^3.
\ee
Taking $M_h\equiv \alpha^3 \, M_{J}$,  the collapse condition becomes
$\alpha_{max}(z) \ge \alpha \ge 1$ where
\be
\alpha_{max}(z) \equiv \left(\frac{M_{\rm upper}(z)}{M_{J}}\right)^{1/3}.
\ee

\begin{figure}[]
\includegraphics[width=0.7\textwidth]{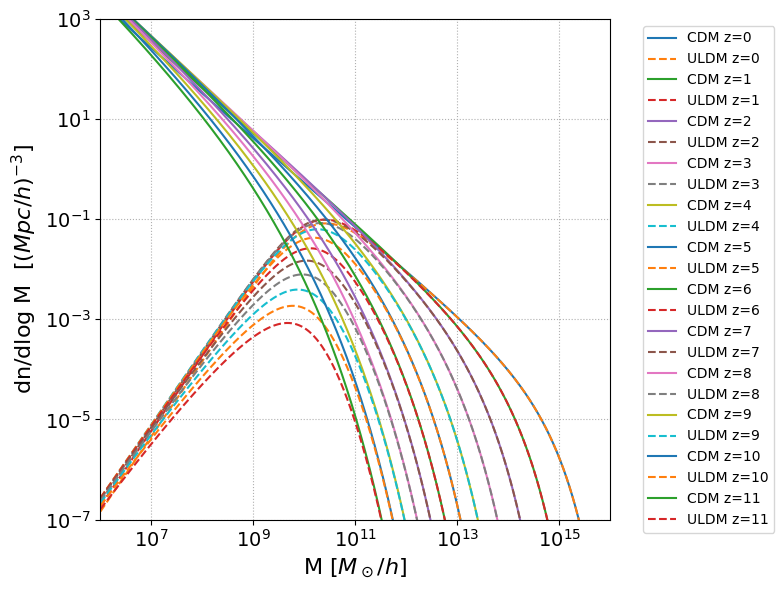}
\caption{ 
Halo mass functions (HMFs) for CDM (solid lines) and ULDM (dashed lines, for $m_{22}=1$) at redshifts $z = 0\sim11$, 
computed using the 
Sheth–Tormen 
form within the Press-Schechter formalism, together with the adopted ULDM suppression factor.
HMFs increase as $z$ decreases, with the highest curves corresponding to $z=0$.
Relative to CDM, ULDM shows a pronounced suppression of low-mass halos arising from the small-scale cutoff in the matter power spectrum. 
}
\label{hmffig}
\end{figure}
This leads to the halo mass as a
function of the parameter $\alpha$
\be
M_h =\frac{4\pi}{3} \rho_{dm}(z) \alpha^3\lambda_J^3(z)=1.204\times 10^8\, \alpha^3 \left(\frac{1+z}{m^2_{22}}\right)^{3/4} M_\odot.
\ee
 To estimate $M_{sol}$
 as a function
 of the halo mass $M_h$, one may use the empirical formula 
for the core-halo relation for ULDM~\cite{PhysRevLett.134.051402}
\be
\label{corehalo}
M_{sol}(M_h)=1.680\times 10^9 m^{-1}_{22}\,\frac{\xi'(z)}{\xi'(7)}\left(\frac{M_h}{10^{11}\,M_\odot}\right)^{1/3}  M_{\odot},
\ee
where $\xi'(z)=\sqrt{1+z}\,\xi(z)^{1/6}$ and $\xi(z)=(18\pi^2+82(\Omega_m(z)-1)-39(\Omega_m(z)-1)^2)/\Omega_m(z)\simeq 177$
for large $z$. When $z>7$ (or even $z>1$), the relative change in $\xi(z)$ can be ignored. Using the halo mass as given above,  the core soliton mass becomes
\be
\label{Msol}
M_{sol}=0.6319\times 10^8 \,\alpha \left(\frac{1+z}{m^2_{22}}\right)^{3/4}M_{\odot}.
\ee
The corresponding half-mass radius of the soliton is given by
\be
\label{rhalf}
r_{1/2}=\frac{2\beta f_0}{3} \frac{\hbar^2}{m^2}\frac{1}{G M_{sol}}= 3.539 
\,\frac{\beta}{ m^{2}_{22}\alpha}  \left(\frac{1+z}{m^2_{22}}\right)^{-3/4}{\rm kpc}
\ee
with $f_0=3.9251$~\cite{Hui:2016ltb}. 
We introduce the phenomenological factor $2\beta/3$ to parametrize a possible reduction of the soliton half-mass radius in the presence of baryons. 
This factor $\beta$ 
is expected to be of order unity, and we set $\beta = 1$ below for simplicity.
To reproduce the observed galactic core sizes, we require $r_{1/2}\simeq {\rm kpc}$ today, which corresponds to $m\simeq10^{-22}{\rm eV}$~\cite{SchiveChiuehBroadhurst2014}.
One finds that, 
compared to the CDM model,
ULDM halos at redshift $z$ can develop relatively compact and high-density cores, particularly in systems hosting massive solitons with large $M_{sol}$.

In this study, the halo mass function (HMF) is required to link the theoretically allowed collapse window to the cosmological abundance of host halos, thereby 
enabling an order-of-magnitude assessment of whether the proposed seed-formation conditions are broadly compatible with the observed SMBH population. We adopt the Press-Schechter (PS) formalism~\cite{1974ApJ...187..425P},
which gives
\be
\frac{dn}{d\ln M} = -\frac{1}{2}\frac{\bar{\rho}_0}{M} f(\nu)\frac{d\ln\sigma^2}{d\ln M},
\ee
where 
$\nu \equiv \frac{\delta_c}{\sigma}
$
with 
the critical overdensity $\delta_c$,
and 
$dn/d\ln M$ is the comoving halo number density per logarithmic mass interval.
For CDM, we adopt the Sheth-Tormen  form for $f(\nu)$~\cite{2001MNRAS.323....1S} for simplicity:
\be
f(\nu) = A \sqrt{\frac{2}{\pi}} \sqrt{q}\nu \Bigl(1 + (q\nu^2)^{-p}\Bigr)
\exp\left[-\frac{q\nu^2}{2}\right]
\ee
with parameters $\{A = 0.3222, p = 0.3, q = 0.707\}$.
Compared to the CDM case, the HMF for ULDM is suppressed at low masses as a result of the quantum-pressure-induced cutoff in the matter power spectrum, and can be approximated as 
\cite{Schive_2016}
\be
\left.\frac{dn}{d\ln M}\right|_{\mathrm{ULDM}}
=
\left.\frac{dn}{d\ln M}\right|_{\mathrm{CDM}}
\left[
1+\left(\frac{M_h}{M_0}\right)^{-1.1}
\right]^{-2.2},
\ee
where 
$M_0 = 1.6 \times 10^{10} m_{22}^{-4/3} M_{\odot}$.
Fig. 1 
shows
HMFs for CDM  and ULDM  at various redshifts $z$, computed within the PS formalism~\cite{murray2013hmfcalconlinetoolcalculating}. Relative to CDM, ULDM shows a pronounced suppression of low-mass halos.

\begin{figure}[]
\includegraphics[width=0.7\textwidth]{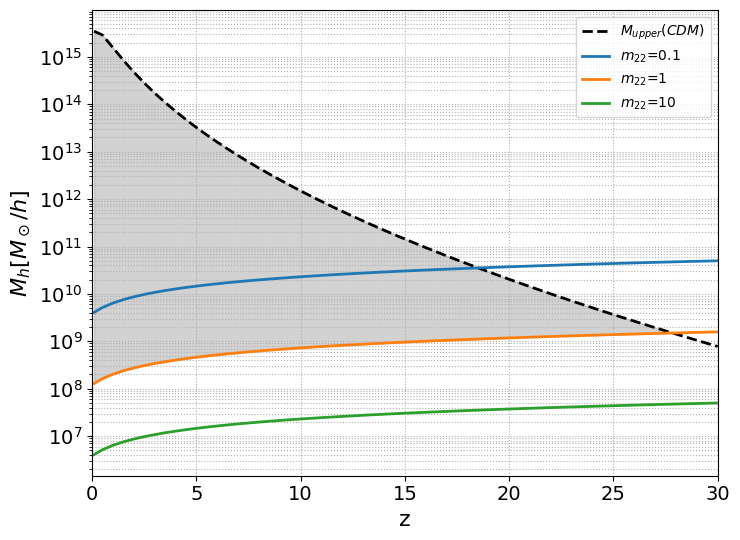}
\caption{ 
The minimum
mass $M_J$ (the thick lines) and
the maximum mass
$M_{upper}$ (the dashed line)
of ULDM halos
as functions of redshift $z$
for $m_{22}=(0.1,1,10)$. 
The halo mass determines the estimated ULDM-core mass and, through the adopted baryonic prescription, the characteristic gas-cloud mass.
For $M_{upper}$, we use
the CDM approximation.
At a given  $z$, 
halos within the interval between $M_J$ and $M_{upper}$ constitute the characteristic population considered in our estimate.
The gray shaded region indicates the allowed halo mass for $m_{22}=1$.
Smaller $m_{22}$ leads to later halo formation.}
\label{Mhaloz}
\end{figure}

To determine $M_{upper}$, one should  integrate the HMF and compare the resulting abundance with observations. 
At a given redshift $z$, we define the maximum halo mass $M_{upper}(z)$ as the mass above which the expected number of halos within the observable Universe over the full sky is exactly unity. To determine this quantity, following
Ref. 
\cite{2012MNRAS.421L..19H}, 
 we  numerically compute the cumulative comoving number density
$n(>M,z)$
from the Sheth-Tormen halo mass function. We then multiply this by the comoving volume element corresponding to a thin redshift slice $dz=0.5$ over the full sky,
to obtain the expected number of halos
$N(>M,z)$. The maximum halo mass
$M_{upper}$
is defined by solving the condition
$ N(>M_{upper}(z),z)=1 $.
This construction provides an estimate of the most massive halo that is  likely to be observed near that redshift. Note that $M_{upper}$ represents a characteristic statistical maximum for the adopted survey volume and redshift interval, rather than an absolute physical upper bound.
For simplicity, we use the CDM halo mass function rather than the ULDM one to compute $M_{upper}$, which is a good approximation for large $M_h$. 
We also assume that the minimum halo mass at  $z$ is set by the Jeans mass $M_J(z)$. With these assumptions, the minimum and maximum 
ULDM
halo masses at a given $z$ can be estimated.
Then, from $M_J$ and 
$M_{upper}$, one can obtain
the ranges of $\alpha$ and $M_{sol}$ 
using Eq. (\ref{corehalo}).
Fig. \ref{Mhaloz} shows $M_J$ and $M_{ upper}$ as functions of $z$ for $m_{22}=(0.1,1,10)$. As $z$ increases, $M_{upper}$ decreases and the allowed range of $M_h$ shrinks.
The soliton mass corresponding to $M_{upper}$ 
from the core-halo relation, namely $M_{sol}(M_h = M_{upper})$, provides 
a characteristic estimate of the largest ULDM core mass expected within the adopted cosmological volume
 at a given redshift. We use this value to estimate the upper end of the gas-cloud and SMBH-seed mass distributions.

\section{Baryonic Collapse and Seed-Mass Scale inside ULDM Solitons}
We now consider baryons embedded within the dark matter soliton.
In realistic collapse scenarios, atomic gas generically develops a finite-density isothermal-like core owing to shocks and efficient radiative cooling~\cite{10.1093/mnras/stad1179}. 
We therefore adopt a pseudo-isothermal profile
\be
 \rho_b(r) =  \frac{\rho_{0}}{1+r^2/r^2_c} 
\ee
as a simple representation of the initial gas cloud
~\cite{trinca2026huntingcosmicgiantsformation,10.1111/j.1365-2966.2011.19168.x},
where $\rho_0$ is the central density and $r_c$ is the core radius of the gas cloud.
For the pseudo-isothermal
profile adopted in this work,
the baryon mass up to radius $r$ is given by
\be
M_b(r)=4 \pi \rho_{0}\, r_c^3 \left(
\frac{r}{r_c}- \arctan \frac{r}{r_c}
\right)
=4 \pi \rho_{0}\, r_c^3 G\left(\frac{r}{r_c}\right),
\ee
where $G(x)\equiv x-\arctan (x)$.
We adopt
this mass profile
for the initial gas distribution 
within the ULDM cores.

The concentration parameter of the pseudo-isothermal gas profile is defined by
$
c_r\equiv r_{\rm vir}/{r_c},
$ and we adopt the representative value $c_r\simeq 30$,
which is 
 motivated by the truncated-isothermal-sphere solution
~\cite{Shapiro_1999}.
Here $r_{\rm vir}$ denotes the characteristic outer radius of the baryonic
cloud confined by the ULDM soliton, rather than the virial radius of the entire
host halo. Since the soliton provides the dominant additional binding within
its central region and its density decreases rapidly outside the half-mass
radius, the natural outer scale of the confined gas cloud is of order
$r_{1/2}$. We therefore parametrize
\beq
r_{\rm vir}=\eta \,r_{1/2},
\eeq
where $\eta$ is an order-unity phenomenological parameter absorbing uncertainties associated with baryonic contraction, angular momentum, and the operational definition of the cloud boundary. The fiducial choice $\eta=1$ is not a calibrated value.
This order-of-magnitude
identification is qualitatively consistent with simulations showing comparable
characteristic sizes for baryonic components and ULDM solitonic cores~\cite{Pozo_2025,Chan_2018}. 
As a fiducial normalization
based on 
the cosmic baryon fraction, we assume that the  baryonic mass confined within $r_{\rm vir}$ is about $20\%$ of the dark-matter mass enclosed within the soliton half-mass radius.
The enclosed baryonic mass is then roughly estimated as
\beq
\label{Mbrvir}
M_b(r_{\rm vir})
\simeq 0.2\,\frac{M_{\rm sol}}{2}
\simeq 4\pi\rho_0r_c^3
G\left(\frac{r_{\rm vir}}{r_c}\right)
=
 4\pi\rho_0r_c^3
G\left(c_r\right).
\eeq

From Eq. (\ref{rhalf}), the baryonic core radius  becomes
\be
\label{rc}
r_c =\eta r_{1/2}/c_r \simeq
118.0\,\eta\times \frac{1}{\alpha m_{22}^2}  \left(\frac{1+z}{m^2_{22}}\right)^{-3/4}{\rm pc}
\ee
for $c_r=30$.
Then, using Eq. (\ref{Mbrvir}), 
one can obtain
\be
M_b(r)\simeq \frac{0.1 M_{sol} G\left(
\frac{r}{r_c}\right) }{G\left(
c_r\right) }
\ee
and the baryon mass
within the core 
\be
\label{Mbrc}
M_b(r_c)
\simeq 
4.764\times 10^4 \alpha  \left(\frac{1+z}{m^2_{22}}\right)^{3/4}{M_\odot}\,,
\ee
using Eq. (\ref{Msol}), 
which characterizes the gas reservoir that may participate in the formation of a massive central object.
The average baryon density enclosed within the radius $r_c$ 
becomes 
\be
\label{rhobrc}
\rho_b(r_c)\simeq \frac{3M_b(r_c)}{4\pi r_c^3} \simeq 
6.929\times 10^{-3} \eta^{-3}\alpha^4 (1+z)^3
\,{\rm M}_{\odot}/{\rm pc}^3.
\ee
For the purpose of an order-of-magnitude estimate, we identify the gas within $r \le r_c$ as the initial baryonic core that may
undergo subsequent collapse via the Jeans instability. 
Therefore, $r_c$ represents the initial size of the baryonic core that undergoes direct collapse. As the collapse proceeds, the core contracts to smaller radii and its baryon density increases beyond the value given in the above equation.
For the adopted baryonic profile with $c_r=30$, the baryonic mass enclosed within 
$r_c$ is estimated to exceed the enclosed ULDM mass by more than an order of magnitude. Thus, once the compact gas core has formed, its subsequent dynamics within  $r_c$ is dominated by baryonic self-gravity.

 The Jeans scale of ULDM determines the minimum halo mass 
 capable of collapsing
 at high redshift, while the empirical soliton-halo relation fixes the soliton mass and dark matter core radius. These, in turn, set the baryonic core density, inflow velocity, shock temperature, and accretion rate.

The mass of the seed black hole  may be expressed as
\beq
\label{MbhMb}
M_{bh}=f_{bh} M_b(r_c),
\eeq
where we adopt a  value of $f_{bh}\simeq O(0.1)$, motivated by the fact that $M_{bh}$ can grow to as much as about half of the quasi-star mass~\cite{hassan2025growthcentralblackholes,Santarelli:2025uck}.
Therefore, 
the characteristic seed mass associated with the adopted parametrization may lie in the range
\beq
\label{bound}
4.764 \, f_{bh}\times 10^4 \left(\frac{1+z}{m^2_{22}}\right)^{3/4}\negthinspace\negthinspace {M_\odot}
\lesssim M_{bh} 
\lesssim
4.764 \, f_{bh}\times 10^4 \alpha_{max}  \left(\frac{1+z}{m^2_{22}}\right)^{3/4}\negthinspace\negthinspace{M_\odot},
\eeq
where $\alpha_{max}$
can be inferred from
$M_{upper}$.
However, we will show that not all seeds within this mass range can form if the temperature of the gas is not sufficiently high.
(See Eqs. (\ref{Mbrcth1}) and (\ref{Mbrcth}) below.)

We now estimate the shock temperature of the core gas cloud as~\cite{Inayoshi_2015}
\begin{equation}\label{shocktem}
T_{core}\simeq \frac{3 \mu m_{\rm H}}{16 k_{\rm B}} v_{\rm in}^2 \simeq
\frac{3 \mu m_{\rm H}}{8 k_{\rm B}} \frac{G M_b(r_c)}{r_c},
\end{equation}
where $v_{in}\simeq \sqrt{2GM_b(r_c)/r_c}$ denotes
the typical gas inflow velocity, and $m_H$ is the hydrogen mass.
Since $r_c\ll r_{1/2}$, the baryonic core is much more compact than the dark-matter soliton core, and therefore $T_{core}$ and $v_{in}$ are governed mainly by the self-gravity of the baryonic gas.
We will use $\mu=1$ as a fiducial normalization in the analytic coefficients. The appropriate value evolves from $\mu\simeq 1.22$ for predominantly neutral primordial gas toward $\mu\simeq 0.59$ for fully ionized gas; hence the numerical shock temperature carries an order-unity composition uncertainty, while the parameter scalings derived below remain unchanged.

This leads to
\be
\label{Tcore}
T_{core}\simeq 78.96 \eta^{-1}\ 
(1+z)^{\frac{3}{2}}
\, \frac{\alpha^{2}}{m_{22}} \, {\rm K},
\ee
which exceeds $10^3\, {\rm K}$ for sufficiently large $\alpha$ at $z>10$ and reaches the atomic-cooling regime only after the stronger threshold conditions below are imposed.
Together with the gas density, this temperature determines whether the collapsing gas initially enters the regime in which $H_2$
cooling is suppressed by collisional dissociation.

From Eq. (\ref{rhobrc}) one can obtain the number density of hydrogen atoms in the initial core 
\be
\label{nH}
n_{H}\simeq 0.2803\eta^{-3}\
\alpha^4 (1+z)^{3}
/ {\rm cm}^3.
\ee
These satisfy the relation
\be
T_{core}/n^{1/2}_{H}=q' \eta^{1/2}/ m_{22} \, {\rm K  cm^{3/2}}
\ee
with $q'=78.96/0.2803^{1/2}\simeq 149.14$. 

For concreteness,
in this paper
we assume the direct collapse scenario in Ref.~\cite{Inayoshi_2012},
where $H_2$ cooling is suppressed when $(T_{core},n_H)$ lies inside the zone of no return.
Although Eq. (\ref{nH}) gives the characteristic density averaged over the initial baryonic core, it provides the correct order of magnitude for the post-shock gas entering the zone of no return. 
We therefore use $(T_{\rm core},n_{\rm H})$ as representative initial conditions for the one-zone thermal-chemical calculations of Ref.~\cite{Inayoshi_2012}.
After  isobaric contraction,
the gas remains on the efficient atomic cooling track at a temperature $T \simeq 8000\, {\rm K}$ and undergoes a nearly isothermal collapse,
avoiding fragmentation and leading to the formation of a supermassive star, which
 can subsequently collapse into a black hole. 
These conditions also imply a favorable hierarchy of physical timescales.
In the direct-collapse pathway of Ref.~\cite{Inayoshi_2012}, the
shock-heated primordial gas first cools efficiently through atomic hydrogen
emission and contracts approximately isobarically toward
$T\simeq 8000\,{\rm K}$.  If the post-shock state lies inside the zone of no
return, newly formed $\mathrm{H}_2$ is collisionally dissociated before
molecular cooling can lower the temperature enough to trigger ordinary
stellar fragmentation.  The gas therefore remains on the atomic-cooling track
and its subsequent evolution is governed primarily by self-gravity, proceeding
on approximately the free-fall timescale
\begin{equation}
 t_{\rm ff}
 =\left(\frac{3\pi}{32G\rho_b}\right)^{1/2}
 \simeq 5.15\times10^{5}
 \left(\frac{n_{\rm H}}{10^{4}\,{\rm cm}^{-3}}\right)^{-1/2}
 {\rm yr}.
 \label{eq:tff_core}
\end{equation}
This is orders of magnitude shorter than the cosmological time 
\begin{equation}
t_{age}(z)
\simeq
1.70\times10^{10}(1+z)^{-3/2}{\rm yr}
\end{equation}
 at
high redshift.  Hence, once the threshold conditions are crossed, the collapse
is dynamically rapid and effectively irreversible: the cloud cannot return to
the low-density, molecular-cooling branch before a massive central object is
assembled. 
In Ref.~\cite{Inayoshi_2012}, the relevant microscopic timescales are already incorporated into the one-zone thermal-chemical calculation and therefore need not be recalculated in our semi-analytic model. 
Accordingly, in our model we calculate the ULDM-dependent dynamical quantities independently, while adopting the subsequent thermal-chemical evolution and the associated cooling timescales from Ref.~\cite{Inayoshi_2012}.
For the large inflow rates relevant to direct collapse, a
supermassive protostellar core can consequently form within
$\sim10^{5}$--$10^{6}\,{\rm yr}$, before ordinary stellar feedback can disrupt
the global inflow~\cite{Inayoshi_2012,Umeda_2016}.

We map our ULDM-generated post-shock conditions onto the one-zone thermal-chemical calculations of
Ref. \cite{Inayoshi_2012}, rather than independently repeating their full reaction-network calculation.
According to those calculations, gas entering this region remains on the atomic-cooling branch because collisional dissociation suppresses the accumulation of $H_2$ and prevents molecular cooling from becoming dynamically important. The gas therefore undergoes an approximately isothermal collapse at $T\simeq 8000 \,{\rm K}$, maintaining the atomic-cooling track up to densities of 
$n_{H}\simeq 10^{16}\,{\rm cm}^{-3}$~\cite{Inayoshi_2012}.

From Eqs. (9) and 
(10) of Ref.~\cite{Inayoshi_2012}, 
the approximate zone-of-no-return boundary, within which collisional dissociation prevents $H_2$ cooling from becoming dynamically important, can be written as
\be
T_{core}  n_H \ge q_l {\rm K / cm^{3}} \quad {\rm and} \quad T_{core}  n_H^{0.1} \ge q_r {\rm K / cm^{0.3}},
\ee
where $q_l=0.56\times 10^8$ and $q_r=0.5\times\sqrt{10}\times 10^4=1.581\times 10^4$.
The two boundary lines meet at $T=(q_r^{10}/q_l)^{1/9} {\rm K}=0.6377\times 10^4 \,{\rm K}$ and $n_H=(q_l/q_r)^{10/9}/cm^3=0.8782\times 10^4/cm^3$.
If the initial values of $T_{core}$ and $n_{H}$ lie within the
zone of no return, 
the gas satisfies the adopted thermal criterion for direct collapse. 
Hence, the direct collapse conditions are satisfied if
\bea
&&T_{core} \ge (q_l (q'\eta^{1/2}/m_{22})^2)^{1/3} {\rm K}=1.076\eta^{1/3}\times 10^4\, {\rm K}/m^{2/3}_{22}\  {\rm and}\nonumber\\
&& T_{core} \ge (q^5_r q'\eta^{1/2}/m_{22})^{1/6} {\rm K}\ =\ 0.7268\eta^{1/12}\times 10^4 \,{\rm K}/m^{1/6}_{22}.
\eea
Since both inequalities must hold, it is sufficient to impose the larger of the two lower bounds on $T_{core}$.
The two lower-bound temperatures become the same if $m_{22}=q' q_l^{2/3}\eta^{1/2}/q_r^{5/3}\equiv a_0=2.192 \eta^{1/2}$.
Therefore, there are two cases.

If
$m_{22}\le a_0 =2.192\eta^{1/2}$, the collapse condition becomes
$
T_{core} \ge (q_l (q'\eta^{1/2}/m_{22})^2)^{1/3} {\rm K}$.
The threshold core temperature is of order $10^4\,{\rm K}$,
which
is high enough to dissociate  $H_2$. 
In this case,
from Eq. (\ref{Tcore}) direct collapse becomes possible if
\be
\alpha^{2}\, \frac{(1+z)^{3/2}}{m^{1/3}_{22}\ } \ \gtrsim\ (q_l {q'}^2)^{1/3}\eta^{4/3}/78.96=136.3 \eta^{4/3}
\ee
or
\be
\label{alphath}
\alpha \ \gtrsim \  \alpha_{th}
= (q_l {q'}^2)^{1/6}\eta^{2/3}/\sqrt{78.96} \times \frac{m^{1/6}_{22}}{(1+z)^{3/4}}=11.67\eta^{2/3} \times \frac{m^{1/6}_{22}}{(1+z)^{3/4}}
\ee
which leads to
\be
\label{Mbrcth1}
M_b(r_c)|_{\alpha=\alpha_{th}}=4.764\times 10^4\times  (q_l {q'}^2)^{1/6}\eta^{2/3}/\sqrt{78.96} \times  m^{-4/3}_{22}\, {\rm M}_\odot=0.5562\times 10^6\times \eta^{2/3}  m^{-4/3}_{22}\, {\rm M}_\odot 
.
\ee
We therefore define the minimum seed mass as
\be
\label{seedmass}
M^{\min}_{bh}\equiv f_{bh}M_b(r_c)|_{\alpha=\alpha_{th}}.
\ee

On the other hand, if
$m_{22} \ge a_0$ (case 2), 
the collapse condition becomes $
T_{core} \ge (q^5_r q'\eta^{1/2}/m_{22})^{1/6} {\rm K}.
$
In this case, direct collapse becomes possible if
\be
\alpha^{2}\, \frac{(1+z)^{3/2}}{m^{5/6}_{22}\ } \ \gtrsim\ (q^5_r q')^{1/6}\eta^{13/12}/78.96=92.04
\eta^{13/12}\ee
or
\be
\label{alphath2}
\alpha \ \gtrsim \  \alpha_{th}
=(q^5_r q')^{1/12}\eta^{13/24}/\sqrt{78.96}\times \frac{m^{5/12}_{22}}{(1+z)^{3/4}}=9.594\eta^{13/24}\times
\frac{m^{5/12}_{22}}{(1+z)^{3/4}}
\ee
which leads to
\be
\label{Mbrcth}
M_b(r_c)|_{\alpha=\alpha_{th}}=4.764\times 10^4\times 
(q^5_r q')^{1/12}/\sqrt{78.96}\times (\eta/m^2_{22})^{13/24}
\, {\rm M}_\odot =0.4571\times 10^6\times (\eta/m^2_{22})^{13/24}
\, {\rm M}_\odot .
\ee

Combining these results with 
the range of minimum seed masses suggested by high-redshift SMBH observations,
$10^4 M_\odot\lesssim M^{\min}_{bh}\lesssim 10^5 M_\odot
$
yields  interesting bounds for the  particle mass of ULDM
\be
3.622~ \eta^{1/2}f_{bh}^{3/4}\lesssim m_{22}\lesssim \min[\,20.37~\eta^{1/2} f_{bh}^{3/4},\, 2.192 \eta^{1/2}],
\ee
for case 1
and
\be
\max\big[\,4.067\,\eta^{1/2}f_{bh}^{12/13}
\negthinspace\negthinspace,\,\, 2.192 \,\eta^{1/2}\big]
\lesssim m_{22}\lesssim 34.07~\eta^{1/2} f_{bh}^{12/13},
\ee
for  case 2. These bounds
are consistent
with the ULDM particle mass $m\simeq O(10^{-22})eV$ commonly inferred from galactic dynamics
for $f_{bh}\simeq O(0.1)$.
For $m\gg 10^{-22}eV$, SMBH seeds are too small in mass, whereas for $m\ll 10^{-22}eV$ their masses are excessively large and the seeds form too late (see Fig. 2).
Interestingly, 
the fiducial value $m\simeq10^{-22}eV$ yields mass scales comparable to the inferred lower seed-mass scale and, under a more speculative extrapolation, the upper end of observed SMBH masses.
(See Fig. 3
and Eq. (\ref{Mbhmax}) below.)
Cores of gas clouds with masses below the thresholds given in Eqs. (\ref{Mbrcth1})
and 
(\ref{Mbrcth})
are less likely to undergo immediate monolithic collapse and may instead fragment into stars or compact stellar systems.
The model may produce an LRD-like phase if, after a massive seed forms, a substantial fraction of the dense gas remains in a compact ionized cocoon.
 
Let us consider case 1 ($m_{22}\le a_0$) for example.
From Eqs. (\ref{alphath}) and  (\ref{rc}), one finds that the initial baryonic core size is smaller than 
\be
r_c|_{\alpha=\alpha_{th}}=10.11~\eta^{1/3} m^{-2/3}_{22} {\rm pc},
\ee
and the radius of the entire gas cloud is smaller than 
\be
r_{\rm vir}|_{\alpha=\alpha_{th}}
=c_r r_c |_{\alpha=\alpha_{th}}\simeq 303.3~ \eta^{1/3}m^{-2/3}_{22} {\rm pc},
\ee
which is broadly comparable to the compact sizes inferred for some LRDs~\cite{Guia_2024},
with effective gas-cloud radii of order
$100 ~{\rm pc}$.

Using $v_{in}\simeq \sqrt{ 2GM_b(r_c)/r_c}$ together with
 Eqs. (\ref{rc}) and (\ref{Mbrc}), the characteristic mass inflow rate
$\dot{M}\simeq 4\pi r_c^2  \rho_b(r_c) v_{in}
\simeq 3\sqrt{2G}M_b(r_c)^{3/2}/r_c^{3/2}$
is given by 
\be
\dot{M}\simeq 2.308 
\eta^{-3/2}\times 10^{-3} \alpha^3 m_{22}^{-3/2} (1+z)^{9/4} {\rm M}_\odot / {\rm yr}
> 3.669~\eta^{1/2}
m_{22}^{-1} {\rm M}_\odot/{\rm yr},
\ee
where we used
the case-$1$ bound $\alpha>\alpha_{th}$
for the last inequality.
The inferred inflow rate is well above the canonical $\sim 0.1 M_\odot/ {\rm yr}$ scale associated with rapidly accreting supermassive protostars. This supports, but does not by itself guarantee, monolithic 
collapse~\cite{Umeda_2016}, providing a possible dynamical pathway to SMBH seeds.
Once SMBH seeds form, ULDM solitons can supply an additional gravitational potential that increases 
the Bondi accretion rate onto a growing black hole seed~\cite{PhysRevLett.134.051402}, thereby enhancing the growth of SMBHs.
Note that we do not consider 
 angular momentum transfer and radiation feedback here,
 for simplicity.
Because the threshold seed mass depends only moderately on $\eta$, we  adopt the fiducial value $\eta = 1$ in Fig. 3.
We also adopt $f_{bh}=0.1$ as a conservative fiducial conversion efficiency, noting that quasi-star calculations can allow substantially larger final BH fractions.

\begin{figure}[]
\includegraphics[width=0.7\textwidth]{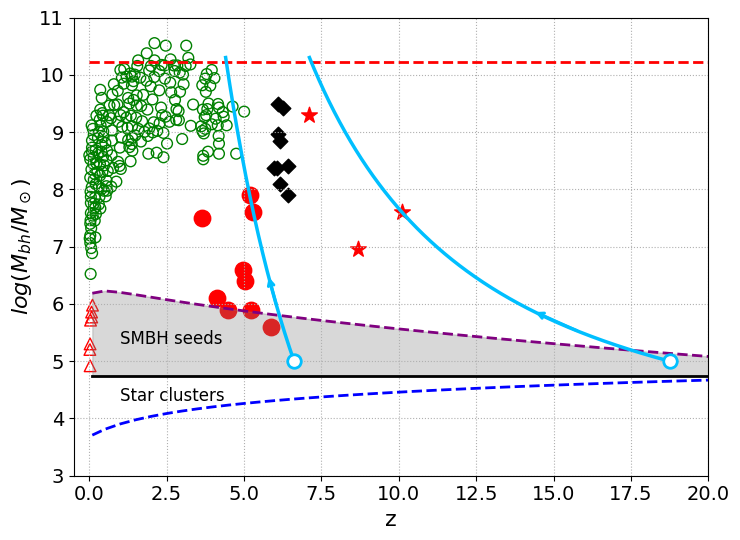}
\caption{ 
Characteristic SMBH mass scales and representative growth tracks as a function of redshift $z$. 
The black horizontal line denotes 
the minimum SMBH seed mass,
$M^{\min}_{bh}$
given in Eq. (\ref{seedmass})
corresponding to
$\alpha_{th}$.
We choose $m=10^{-22}eV$,
$c_r=30$,
 $\eta=1$ and $f_{bh}=0.1$ here.
The blue dashed line
and the purple dashed
line denote
the possible SMBH seed masses in Eq. (\ref{bound})
for $\alpha=1$ and
$\alpha=\alpha_{max}$, respectively.
However, objects located between the black horizontal line and the blue line 
do not satisfy the adopted thermal threshold and are therefore more likely to fragment than to undergo immediate monolithic collapse. 
These systems may instead  form compact stellar systems.
According to Eq. (\ref{Mbrcth1}) in our model, 
only gas clouds in the gray region satisfy the adopted semi-analytic thermal criterion and are classified as direct-collapse candidates for SMBH seed formation.
The dashed horizontal red line denotes an illustrative upper mass scale obtained from Eq. (\ref{Mbhmax}).
Dots represent the observed masses of SMBHs from various observations~\cite{Dewangan2008ApJ,Vestergaard_2009,Willott_2010,Mortlock_2011,Larson_2023,natarajan2023detectionovermassiveblackhole}.
The red triangles denote XMM-Newton observations~\cite{Dewangan2008ApJ}, the green circles represent LBQS observations~\cite{Vestergaard_2009}, the black diamonds denote CFHQS
observations~\cite{Willott_2010}. The red stars denote quasars with $z>7$~\cite{Mortlock_2011,Larson_2023,natarajan2023detectionovermassiveblackhole}, while the red star at $z=10.1$
represents 
the quasar in the galaxy UHZ1.
The red disks denote
the candidate SMBHs in LRDs~\cite{Rusakov2026LittleRedDots}.
The blue lines denote
example growth trajectories
of initial SMBH seeds
with $M_i=10^5 M_\odot$ 
and a radiative efficiency of $\epsilon=0.1$. 
}
\label{omegafig}
\end{figure}

In our model, SMBH seeds can form
at  high redshifts, $z>10$, possibly
alleviating tensions in standard black-hole formation scenarios.
Assuming Eddington-limited accretion with a radiative efficiency of $\epsilon=0.1$, the black hole mass evolves as~\cite{moffat2020supermassiveblackholeaccretion}
\be
M(z_f)=M_i \exp\left[\frac{t(z_f)-t(z_i)}{t_{\rm Sal}}\right],
\ee
where $t_{\rm Sal}\simeq 45\,{\rm Myr}$
is the Salpeter e-folding timescale.
 For example, for 
 the quasar in the galaxy UHZ1  with  mass $M(z_f)=4\times10^7M_\odot$ at $z_f=10.1$~\cite{natarajan2023detectionovermassiveblackhole}, starting from an initial seed mass $M_i=10^5\,M_\odot$, the required growth factor is $M(z_f)/M_i=400$. This implies  $t(z_f)-t(z_i)\simeq 2.7\times10^8{\rm yr}$, which corresponds to an initial redshift $z_i\simeq 19$. (See Fig. 3.) Therefore,
 a heavy seed of mass $10^5M_\odot$ in our scenario has sufficient cosmic time to grow  to $\sim 4\times10^7 M_\odot$ by $z\simeq 10$
 under continuous Eddington-limited accretion with $\epsilon=0.1$ and a duty cycle of unity, demonstrating that early SMBH growth is feasible without invoking super-Eddington accretion.
 Fig. 3 illustrates that the characteristic mass scales and representative growth tracks of the model are broadly compatible with several observed quasars and candidate SMBHs in LRDs.
UHZ1 at $z = 10.1$ is among the earliest and therefore most constraining objects in our sample. Its growth from a $10^5 M_{\odot}$ seed formed at $z_i\simeq 19$ 
is compatible with continuous Eddington-limited accretion for $\epsilon = 0.1$. Objects observed at lower redshifts have more cosmic time available for growth. Their positions below the corresponding continuous Eddington-growth envelope therefore indicate that super-Eddington accretion is not required by the available growth time, although the detailed evolution depends on the duty cycle, radiative efficiency, and seed-formation epoch. For comparison, Fig. 3 also shows a representative Eddington-limited growth trajectory for the SMBH candidates in LRDs.
Our model allows SMBH seeds with $M_{bh}\gtrsim 10^5 M_\odot$ to form at $z>7$, potentially producing an LRD-like phase.

When all the parameters are explicitly retained, the minimum black-hole seed mass
for  case 1 can be written as
\beq
\label{Mbhmin2}
M_{bh}^{\min}\simeq 0.5562\times10^{6}f_{bh}\mu^{-1/6}\eta^{2/3}\left[\frac{30G(30)}{c_rG(c_r)}\right]^{2/3}m_{22}^{-4/3}M_\odot,
\eeq
which depends only weakly on the mean molecular weight, but more appreciably on the cloud size and gas concentration. In particular, for $c_r\gg1$, $M_{bh}^{\min}\propto\eta^{2/3}c_r^{-4/3}$.
Within the plausible ranges of $\eta$ and $c_r$, the minimum seed mass varies only within about one order of magnitude.
(See Fig. 4.) Thus, although the normalization of $M_{bh}^{\min}$ depends on the assumed gas distribution, the order of magnitude of the predicted seed mass remains unchanged.

The bounds
in Eq. (\ref{bound})
apply only to black hole
seeds, not to all SMBHs
grown from the seeds
in our scenario.
To estimate the maximum mass of SMBHs,
we need to determine
the physically allowed maximum mass of
gas clouds in an
extremely massive ULDM core.
For this purpose, we consider 
another well-known upper bound 
on the soliton mass,
independent of $z$,
derived from boson star theory~\cite{Liebling:2012fv}.
The maximum gravitationally stable mass of an isolated ground-state configuration
of ULDM is given by
\begin{equation}
\label{Mmax}
M_{{\max}} \simeq 0.633\frac{M_{\mathrm{Pl}}^{2}}{m}\simeq
\frac{8.5 \times 10^{11}}{m_{22}} M_{\odot},
\end{equation}
where $M_{\mathrm{Pl}}$ is the Planck mass.
Solitons near or above this stability limit are relativistically unstable and may migrate to a lower-mass configuration, or collapse to a black hole depending on their dynamical state~\cite{Liebling:2012fv}.
Note that $M_{\max}$ above,
which is derived from
boson star dynamics,
is usually much larger than the soliton mass bound
$M_{sol}(M_h=M_{upper})$,
which is obtained from a statistical analysis.
These two mass scales
therefore have different physical meanings and origins.
In principle, repeated mergers could increase the mass of a central solitonic configuration up to $M_{max}$~\cite{Hui:2016ltb,Lee:2023krm}
in extreme situations.
This may, in turn, provide a physical upper bound on the masses of gas clouds within the ULDM solitons and on the maximum masses of black holes grown by consuming these baryonic gas clouds.
Under these assumptions, this provides a speculative characteristic scale for the most massive SMBHs, rather than merely that of the seeds,
\beq
\label{Mbhmax}
M_{bh}^{\max}\equiv 0.2f_{bh}M_{\max}
\simeq
\frac{1.7 \times 10^{11} f_{bh}}{m_{22}} M_{\odot},
\eeq
which is proportional to the $total$ baryonic mass within the soliton, whereas $M_{bh}$
in Eq. (\ref{MbhMb}) is proportional  to the baryonic mass enclosed only within $r_c$.
The dashed
horizontal red line
in Fig. 3 denotes 
$M_{bh}^{\max}$.
Interestingly, using the observational estimate $M_{bh}^{\max}\simeq
10^{10}M_\odot$~\cite{Inayoshi2020},
one may again obtain
$m\simeq O(10^{-22})eV$
for $f_{bh}=O(0.1)$.
Note that we did not
rely on the core-halo relation to calculate 
$M_{\max}$ and $M_{bh}^{\max}$.

\begin{figure}[]
\includegraphics[width=0.7\textwidth]{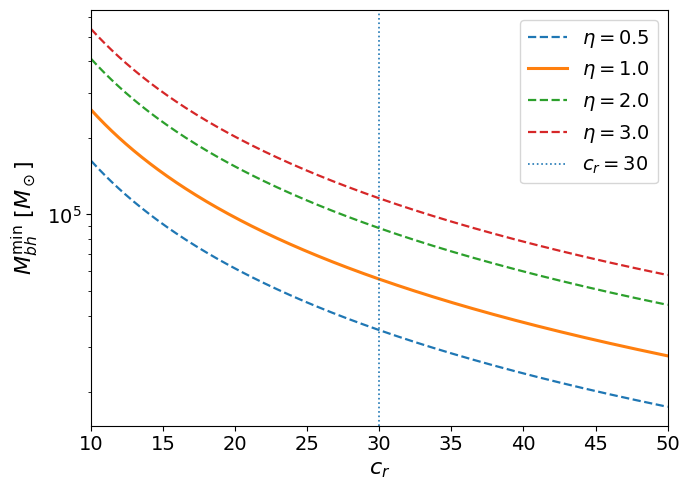}
\caption{  Dependence of the minimum black-hole seed mass, $M_{bh}^{\min}$
in Eq. (\ref{Mbhmin2}), on the gas concentration parameter $c_r$ for several values of the cloud-size parameter $\eta$. We set $\mu=1=m_{22}$ and $f_{bh}=0.1$. The solid curve denotes the fiducial case $\eta=1$, while the dashed curves show the dependence on $\eta$. The vertical dotted line indicates the fiducial value $c_r=30$.
}
\label{crfig}
\end{figure}

\section{Discussion}
Observations of LRDs indicate that black holes with masses
$M_{\rm bh}\gtrsim 10^5 M_\odot$
already exist at $z\gtrsim 5$, and our scenario provides a  physical interpretation of their origin. In this work,
we find that this characteristic mass scale can emerge from the adopted ULDM scaling relations: when the baryonic core inside a soliton satisfies the temperature threshold, the enclosed gas mass reaches
$M_b(r_c)\big|_{\alpha=\alpha_{\rm th}}\sim10^6M_\odot$,
and a fraction $f_{\rm bh}\sim0.1$ of this mass is converted into a black hole, yielding a seed of order $10^5M_\odot$ 
for fiducial values of the phenomenological parameters.
A key outcome is that, within the adopted core-halo relation and baryonic parametrization, the characteristic seed mass follows from the ULDM mass and length scales rather than being inserted independently.

The compact configuration set by the soliton permits the baryonic core to become self-gravitating; the resulting inflow can then produce characteristic shock temperatures approaching $10^4 \,{\rm K}$ in the threshold region,
suppressing molecular hydrogen cooling and preventing fragmentation, consistent with the compact, hot, and ionized environments inferred for LRDs. 
Within the adopted thermal criterion, clouds below this threshold are less likely to undergo immediate monolithic collapse and are instead expected to collapse into dense star clusters, possibly accounting for the observed coexistence of AGN-like and young stellar-dominated LRDs~\cite{Carranza_Escudero_2025}, and 
potentially triggering the early formation of galaxies within a single physical framework.
Furthermore, because the relevant Jeans scale and the soliton-halo relation favor the formation of dense cores at high redshift, 
our scenario permits favorable seed-formation conditions in sufficiently massive and rare halos at $z \ge  10$,
potentially accounting for
the 
high-redshift range where LRDs are observed by JWST. 
Finally, continuous Eddington-limited accretion with a duty cycle near unity provides an upper-envelope history capable of connecting a $10^5 M_\odot$ seed to some observed masses by $z\sim 10$.

This mechanism has several  implications
for early structure formation. 
It suggests a possible connection
between the small-scale physics of ULDM and the early growth of supermassive black holes, implying 
that the existence and typical masses of high-redshift SMBHs may carry indirect information about ULDM particle properties. At the same time, 
the scenario yields testable characteristic ranges of host-halo masses and thermodynamic conditions for early black hole formation, which can be compared with upcoming observations of the high-redshift Universe.
Photometric JWST observations have reported candidate galaxies extending to $z\sim 20$~\cite{Yan_2022}. Theoretical halo mass function calculations and cosmological simulations of ULDM further show that sufficiently massive perturbations above the Jeans or cutoff scale can undergo nonlinear collapse at very high redshifts, potentially as early as $z\sim 20$~\cite{Lidz_2018,May_2023}.
 
The effects of angular momentum and radiative feedback are not included in the present semi-analytic treatment, although the solitonic potential is expected to promote rapid infall. 
While our model is broadly compatible with several properties inferred for LRDs and high-redshift SMBHs, including their characteristic mass scales, formation epochs, and association with hot, ionized gas, a more quantitative comparison with their number densities and host-halo properties is still needed.
While the Press-Schechter approach provides a useful estimate, more accurate ULDM halo mass functions derived from simulations should be explored in future work. Addressing these issues will require high-resolution simulations 
\cite {10.1093/mnras/stad1179} and improved observational constraints, and
the present work should therefore be regarded as a starting point for more detailed studies linking the small-scale physics of ULDM to the observed population of high-redshift SMBHs.

\subsection*{Acknowledgments}
DB was supported in part by 
Basic Science Research Program through NRF, funded by the Ministry of Education (2018R1A6A1A06024977).
JW was supported in part by the National Supercomputing Center with supercomputing resources including technical support (KSC-2023-CRE-0250).


\end{document}